\documentclass[twoside,12pt]{article}
\usepackage{CJK}
\usepackage{indentfirst}
\usepackage{bm}
\usepackage{verbatim}
\usepackage{graphicx}
\usepackage{multicol}

\begin{document}
\begin{CJK*}{GBK}{song}
\newcommand{\song}{\CJKfamily{song}}
\newcommand{\hei}{\CJKfamily{hei}}
\newcommand{\fs}{\CJKfamily{fs}}
\newcommand{\kai}{\CJKfamily{kai}}
\def\thefootnote{\fnsymbol{footnote}}
\begin{center}
\Large\hei  Squeezing of light field in a dissipative Jaynes-Cummings model $^{*}$%% 论文题目
\end{center}

%\footnotetext{\hspace*{-.45cm}\footnotesize $^*$Project supported by the Science and Technology Plan of Hunan Province, China(Grant No 2010FJ3148), the National Natural Science Foundation of China (Grant No 11374096) and the Doctoral Science Foundation of Hunan Normal University, China.}
\footnotetext{\hspace*{-.45cm}\footnotesize $^\dag$Corresponding author. E-mail:zhmzc1997@126.com, tel:13807314064}

\begin{center}
\rm Hong-Mei Zou $^{\dagger}$, \ Mao-Fa Fang
\end{center}

\begin{center}
\begin{footnotesize} \rm
Key Laboratory of Low-dimensional Quantum Structures and Quantum Control of Ministry of Education, College of Physics and Information Science, Hunan Normal University, Changsha, 410081, China\  \\   %%%% 地址 a)

%%% 更多地址依次往下延续
\end{footnotesize}
\end{center}

\begin{center}
\footnotesize (Received XX; revised manuscript received X XX)
          %% (Received 日 月 年; revised manuscript received 日 月 年)
\end{center}

\vspace*{2mm}

\begin{center}
\begin{minipage}{15.5cm}
\parindent 20pt\footnotesize

Based on the time-convolutionless master-equation approach, we investigate squeezing of light field in a dissipative Jaynes-Cummings model. The results show that squeezing light can be generated when the atom transits to a ground state from an excited state, and then a collapse-revival phenomenon will occur in the squeezing of light field due to atom-cavity coupling. Enhancing the atom-cavity coupling can increase the frequency of the collapse-revival of squeezing. The stronger the non-Markovian effect is, the more obvious the collapse-revival phenomenon is. The oscillatory frequency of the squeezing is dependents on the resonant frequency of the atom-cavity.

\end{minipage}
\end{center}

\begin{center}
\begin{minipage}{15.5cm}
\begin{minipage}[t]{2.3cm}{\hei Keywords: }\end{minipage}
\begin{minipage}[t]{13.1cm} squeezing of light field, Jaynes-Cummings model, non-Markovian environment
\end{minipage}\par\vglue8pt
{\bf PACS: }03.65.Yz, 03.67.-a, 42.50.Lc, 42.50.Pq
\end{minipage}
\end{center}

\section{\fs Introduction}  %%% 节标题 1
Squeezed light, as one of quantum properties of light field, has been widely investigated since it was observed in a groundbreaking experiment of an atomic vapor of Sodium atoms\cite{Slusher55}. In the past 30 years, many significant progresses have been acquired in experimental and theoretical researches on squeezed light so that squeezed light has a wide range of applications in quantum information processing and quantum metrology\cite{Buzek2004,Braunstein77,Andersen4}.

The basic idea of squeezing can be understood by considering the quantum harmonic oscillator. For a vacuum state, the variance of the position and momentum observables equals $\frac{1}{2}\hbar$, so we say that a state of a single harmonic oscillator exhibits squeezing if the variance of the position or momentum observables is below $\frac{1}{2}\hbar$. In order to obey Heisenberg's uncertainty relation\cite{Heisenberg,Robertson}, both position and momentum observables cannot be simultaneously squeezed. In the quantum optics theory, the Hilbert space associated with a mode of the electromagnetic field is isomorphic to that of the mechanical harmonic oscillator, thus the position and momentum observables in the harmonic oscillator correspond to the electric field magnitudes measured at specific phases, that is, the position and momentum operators of the mechanical harmonic oscillator can be expressed by the creation and annihilation operators of a quantum field, i.e. $\hat{x}=\sqrt{\frac{\hbar}{2\omega}}(\hat{a}+\hat{a}^{\dag})$, $\hat{p}=-i\sqrt{\frac{\hbar\omega}{2}}(\hat{a}-\hat{a}^{\dag})$ and $[\hat{x},\hat{p}]=i\hbar$.
Accordingly, the field measurements in an electromagnetic wave are affected by quantum uncertainties. For the coherent and vacuum states, this uncertainty is $(\triangle x)^{2}\cdot(\triangle p)^{2}=\frac{1}{4}\hbar^{2}$, called the standard quantum limit\cite{Loudon34,Lvovsky1401,Andersen1511}, where $(\triangle X)^{2}=\langle X^{2}\rangle-\langle X\rangle^{2}$ ($X=x,p$). But squeezed states of light field exhibit uncertainties below the standard quantum limit\cite{PengJS}. This means that, when we turn on the squeezed light, we see less noise than no light at all. Hence, squeezed light is an interesting physical reality and has a variety of applications, such as precision measurements of distances, gravitational wave detectors, universal quantum computing, dense coding and quantum key distribution, and so on. Though the original definition of squeezed states refers to the squeezing of the quadrature amplitudes, squeezing of other quantities has also been studied including photon number squeezing and two-mode squeezing\cite{Caves31,Reid60}. And, there exist many physical processes that can be employed to prepare single- and two-mode states of the electromagnetic field. For examples, single-mode squeezing lights and two-mode squeezing lights can be generated via parametric down-conversion\cite{Boyd2003,Kim73,LAWu57}, in atomic ensembles\cite{Slusher55,McCormick32,Barreiro84} and in optical fibers\cite{Heersink30,Heersink33}.

Study of squeezed light is a hot issue in quantum optics. Recently, a lot of new works are reported in the research area of squeezed light. The authors in Ref.\cite{Mi Zhang1507} investigated the spatial distribution of quantum fluctuations in a squeezed vacuum field. The authors in Ref.\cite{Bhattacherjee1411} discussed influence of virtual photon process on the generation of squeezed light from atoms in an optical cavity. The authors in Ref.\cite{Berchera92} considered in detail a system of two interferometers aimed to the detection of extremely faint phase-fluctuations. The authors in Ref.\cite{Lucivero1509} researched theoretically the squeezing spectrum and second-order correlation function of the output light for an optomechanical system, and so on. In this present work, we investigate squeezing of light field in a dissipative Jaynes-Cummings model and  analyze influences of initial atomic states, atom-cavity couplings, non-Markovian effects and resonant frequencies on the squeezing. The results show that a collapse-revival phenomenon will occur in the squeezing of light field. Enhancing the atom-cavity coupling can increase the frequency of the collapse-revival. The stronger the non-Markovian effect is, the more obvious the collapse-revival phenomenon is. The oscillatory frequency of squeezing is dependents on the resonant frequency of the atom-cavity.

The outline of the paper is the following. In Section 2, we introduce a dissipative Jaynes-Cummings model. In Section 3, we review the concept of squeezing of light field. Results and Discussions are given in Section 4. Finally, we sum up the report in Section 5.

\section{\fs Dissipative Jaynes-Cummings model}
We consider a composite system of a two-level atom interacting with a cavity which is coupled to a bosonic environment\cite{Scala1,ZouHM2}. The Hamiltonian is written as ($\hbar=1$)
\begin{eqnarray}\label{EB01}
H=H_{JC}+H_{r}+H_{I}
\end{eqnarray}
here
\begin{eqnarray}\label{EB02}
H_{JC}&=&\frac{1}{2}\omega_{0}\sigma_{z}+\omega_{0}a^{\dag}a+\Omega(a\sigma_{+}+a^{\dag}\sigma_{-}),\nonumber\\
H_{r}&=&\sum_{k}\omega_{k}b_{k}^{\dag}b_{k},\nonumber\\
H_{I}&=&(a^{\dag}+a)\sum_{k}g_{k}(b_{k}^{\dag}+b_{k}),
\end{eqnarray}
where $\sigma_{+}=|e\rangle\langle g|$ and $\sigma_{-}=|g\rangle\langle e|$ are the raising and lowering operators of the qubit, and $\sigma_{z}=|e\rangle\langle e|-|g\rangle\langle g|$ is a Pauli operator for the qubit with transition frequency $\omega_{0}$\cite{Jaynes}. $a^{\dag}$ and $a$ are the creation and annihilation operators of the cavity field. $b_{k}^{\dag}$ and $b_{k}$ are the creation and annihilation operators of the environment. $\Omega$ is the coupling between the atom-cavity and $g_{k}$ is the coupling between the cavity-environment.

Let us suppose that there is one initial excitation in the atom-cavity system and the environment is at zero temperature. Neglecting the atomic spontaneous emission and the Lamb shifts, in the dressed-state basis $\{|E_{1+}\rangle, |E_{1-}\rangle, |E_{0}\rangle\}$, using the second order of the time convolutionless(TCL) expansion\cite{Breuer1}, the non-Markovian master equation for the density operator $\Lambda(t)$ is
\begin{eqnarray}\label{EB05}
\dot{\Lambda}(t)&=&-i[H_{JC},\Lambda(t)]+\gamma(\omega_{0}+\Omega,t)(\frac{1}{2}|E_{0}\rangle\langle E_{1+}|\Lambda(t)|E_{1+}\rangle\langle E_{0}|-\frac{1}{4}\{|E_{1+}\rangle\langle E_{1+}|,\Lambda(t)\})\nonumber\\
&+&\gamma(\omega_{0}-\Omega,t)(\frac{1}{2}|E_{0}\rangle\langle E_{1-}|\Lambda(t)|E_{1-}\rangle\langle E_{0}|-\frac{1}{4}\{|E_{1-}\rangle\langle E_{1-}|,\Lambda(t)\}),
\end{eqnarray}
where $|E_{1\pm}\rangle=\frac{1}{\sqrt{2}}(|1g\rangle\pm|0e\rangle)$ are the eigenstates of $H_{JC}$ with energy $\frac{\omega_{0}}{2}\pm\Omega$ and $|E_{0}\rangle=|0g\rangle$ is the ground state with energy $-\frac{\omega_{0}}{2}$. The timedependent decay rates for $|E_{1-}\rangle$ and $|E_{1+}\rangle$ are $\gamma(\omega_{0}-\Omega,t)$ and $\gamma(\omega_{0}+\Omega,t)$ respectively.

We take a Lorentzian  spectral density of the environment, i.e.
\begin{equation}\label{EB06}
J(\omega)=\frac{1}{2\pi}\frac{\gamma_{0}\lambda^{2}}{(\omega_{1}-\omega)^{2}+\lambda^{2}},
\end{equation}
where $\gamma_{0}$ is related to the relaxation time scale $\tau_{S}$ by $\tau_{S}$=$\gamma_{0}^{-1}$ and $\lambda$ defines the spectral width of the coupling which is connected to the reservoir correlation time $\tau_{R}$ by $\tau_{R}$=$\lambda^{-1}$. If $\lambda>2\gamma_{0}$, the relaxation time is greater than the reservoir correlation time and the dynamical evolution of the system is essentially Markovian. For $\lambda<2\gamma_{0}$, the reservoir correlation time is greater than or of the same order as the relaxation time and non-Markovian effects become relevant\cite{Bellomo1,ZouHM3,ZouHM4}. When the spectrum is peaked on the frequency of the state $|E_{1-}\rangle$, i.e. $\omega_{1}=\omega_{0}-\Omega$, the decay rates for the two dressed states $|E_{1\pm}\rangle$ are respectively expressed as\cite{Scala1} $\gamma(\omega_{0}-\Omega,t)=\gamma_{0}(1-e^{-\lambda t})$ and $\gamma(\omega_{0}+\Omega,t)=\frac{\gamma_{0}\lambda^{2}}{4\Omega^{2}+\lambda^{2}}\{1+[\frac{2\Omega}{\lambda}sin2\Omega t-cos2\Omega t]e^{-\lambda t}\}$.

If the initial state of the atom-cavity is
\begin{eqnarray}\label{EB09}
\Lambda(0)=\left(
            \begin{array}{cccc}
              \Lambda_{11}(0)&\Lambda_{12}(0)&\Lambda_{13}(0)\\
              \Lambda_{21}(0)&\Lambda_{22}(0)&\Lambda_{23}(0)\\
              \Lambda_{31}(0)&\Lambda_{32}(0)&\Lambda_{33}(0)\\
            \end{array}
          \right),
\end{eqnarray}
we can acquire the matrix elements at all times from Equation ~(\ref{EB05})
\begin{eqnarray}\label{EB10}
     \Lambda_{11}(t)&=&c_{11}^{11}\Lambda_{11}(0),      \Lambda_{12}(t)=c_{12}^{12}\Lambda_{12}(0),      \Lambda_{13}(t)=c_{13}^{13}\Lambda_{13}(0),\nonumber\\
     \Lambda_{22}(t)&=&c_{22}^{22}\Lambda_{22}(0),      \Lambda_{23}(t)=c_{23}^{23}\Lambda_{23}(0),\nonumber\\
     \Lambda_{33}(t)&=&c_{33}^{11}\Lambda_{11}(0)+c_{33}^{22}\Lambda_{22}(0)+c_{33}^{33}\Lambda_{33}(0),
\end{eqnarray}
here
\begin{eqnarray}\label{EB11}
      c_{11}^{11}&=&e^{-f_{2}},   c_{12}^{12}=e^{-2i\Omega t}e^{-\frac{1}{2}(f_{2}+f_{1})}, c_{13}^{13}=e^{-i(\omega_{0}+\Omega)t}e^{-\frac{1}{2}f_{2}},\nonumber\\
      c_{22}^{22}&=&e^{-f_{1}},   c_{23}^{23}=e^{-i(\omega_{0}-\Omega)t}e^{-\frac{1}{2}f_{1}},\nonumber\\
      c_{33}^{11}&=&1-c_{11}^{11},   c_{33}^{22}=1-c_{22}^{22},      c_{33}^{33}=1,
\end{eqnarray}
where
\begin{eqnarray}\label{EB12}
f_{1}&=&\frac{1}{2}\int_{0}^{t}\gamma(\omega_{0}-\Omega,t')dt'=\frac{1}{2}(\gamma_{0}t+\frac{\gamma_{0}}{\lambda}(e^{-\lambda t}-1))
\end{eqnarray}
and
\begin{eqnarray}\label{EB12}
f_{2}&=&\frac{1}{2}\int_{0}^{t}\gamma(\omega_{0}+\Omega,t')dt'\nonumber\\
&=&\frac{1}{2}\frac{\gamma_{0}\lambda^{2}}{4\Omega^{2}+\lambda^{2}}[t-\frac{4\Omega e^{-\lambda t} sin(2\Omega t)}{4\Omega^{2}+\lambda^{2}}+\frac{(\lambda^{2}-4\Omega^{2})(e^{-\lambda t}cos(2\Omega t)-1)}{\lambda(4\Omega^{2}+\lambda^{2})}].
\end{eqnarray}

\section{Squeezing of light field}  %%%  3.
To analyze the squeezing property of light field, we review two quadrature operators $X_{1}$ and $X_{2}$ having a phase difference $\frac{\pi}{2}$\cite{PengJS}, i.e.
\begin{eqnarray}\label{EB14}
X_{1}&=&\frac{1}{2}(a+a^{+}),\nonumber\\
X_{2}&=&\frac{1}{2i}(a-a^{+})
\end{eqnarray}
where $[X_{1},X_{2}]=\frac{i}{2}$. The Heisenberg uncertainty relationis given by
\begin{eqnarray}\label{EB15}
(\triangle X_{1})^{2}\cdot(\triangle X_{2})^{2}& \geq &\frac{1}{16}
\end{eqnarray}
where $(\triangle X_{j})^{2}=\langle X_{j}^{2}\rangle-\langle X_{j}\rangle^{2}$ (j=1,2). Consequently,
the fluctuation in the amplitudes $X_{j}$ of the quadrature operators is said to be squeezed if $X_{j}$ satisfies the condition
\begin{eqnarray}\label{EB16}
(\triangle X_{j})^{2}& <&\frac{1}{4}
\end{eqnarray}
or
\begin{eqnarray}\label{EB17}
F_{j}&=&(\triangle X_{j})^{2}-\frac{1}{4}<0
\end{eqnarray}

If we assume that after time t the density matrix of the light field has the following form
\begin{eqnarray}\label{EB18}
\rho_{f}(t)&=&\left(
            \begin{array}{cc}
              \rho_{11}(t) & \rho_{12}(t)\\
              \rho_{21}(t)& \rho_{22}(t)\\
                          \end{array}
          \right),
\end{eqnarray}
we can get $\langle X_{j}\rangle= Tr(X_{j}\rho_{f}(t))$ and $\langle X_{j}^{2}\rangle= Tr(X_{j}^{2}\rho_{f}(t))$. Then we insert $\langle X_{j}\rangle$ and $\langle X_{j}^{2}\rangle$ into Equation ~(\ref{EB16}) or Equation ~(\ref{EB17}) and can analyze numerically the squeezing properties of the light field.

\section{\fs Results and Discussions}
We set that the initial state of the atom-cavity is
\begin{eqnarray}\label{EB15}
     \begin{array}{cccc}
      |\psi(0)\rangle=(cos(\frac{\theta}{2})|e\rangle+e^{i\varphi}sin(\frac{\theta}{2})|g\rangle)_{A}|0\rangle_{f}\\
     \end{array},
\end{eqnarray}
where $A$ indicates the atom, $f$ expresses the cavity field. $\theta$ is an amplitude parameter and $\varphi$ is a phase parameter. We can obtain the reduced density matrix $\rho_{f}(t)$ by using Equation ~(\ref{EB10}). In the following, we analyse the influences of the initially atomic state, the atom-cavity coupling, the resonant frequency of the atom-cavity and the non-Markovian effect on the squeezing of light field.

\begin{center}
\includegraphics[width=16cm,height=6cm]{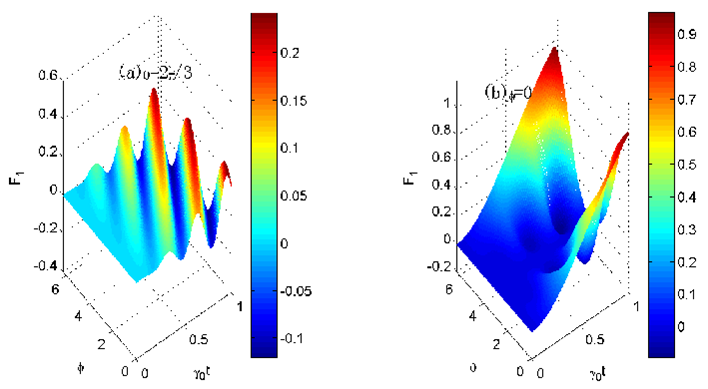}
\parbox{16cm}{\small{\bf Figure 1.}
(Color online)The influence of the initially atomic state on the squeezing of light field in the Markovian regime($\lambda=5\gamma_{0}$). (a)The dynamics revolution of the squeezing factor $F_{1}$ as functions of $\varphi$ and $\gamma_{0}t$ when $\theta=\frac{2\pi}{3}$; (b)The dynamics revolution of the squeezing factor $F_{1}$ as functions of $\theta$ and $\gamma_{0}t$ when $\varphi=0$. Other parameters are $\Omega=\gamma_{0}$ and $\omega_{0}=10\gamma_{0}$.}
\end{center}

In Figure 1, we describe the influence of the initial atomic states on the squeezing of light field in the Markovian regime($\lambda=5\gamma_{0}$). Figure 1(a) shows the squeezing factor $F_{1}$ as functions of $\varphi$ and $\gamma_{0}t$ when $\theta=\frac{2\pi}{3}$. From Figure 1(a), we see that the $F_{1}$ is obvious $t$ dependent but is $\varphi$ independent. For a certain value of $\varphi$, the $F_{1}$ varies to a negative value from zero and then oscillates when time $t$ increases, namely, there is not the squeezing in the initial state but the atomic transition can generate squeezing light, and then a collapse-revival phenomenon will occur in the squeezing of light field due to the atom-cavity coupling. For different $\varphi$, the $F_{1}$ has similarly oscillatory behavior as time $t$ increases. Figure 1(b) depicts the squeezing factor $F_{1}$ as functions of $\theta$ and $\gamma_{0}t$ when $\varphi=0$. From Figure 1(b), we find that, the $F_{1}$ is not only dependent on $t$ but also dependent on $\theta$. For different $\theta$, there are the diverse evolution curves of $F_{1}$, for examples, there is no any squeezing phenomena when $\theta=\pi$ while the $F_{1}$ is squeezed for some certain values of $\theta$.

\begin{center}
\includegraphics[width=16cm,height=10cm]{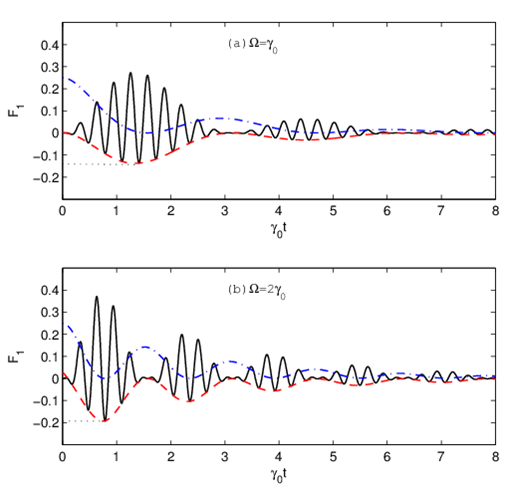}
\parbox{16cm}{\small{\bf Figure 2.}
(Color online)The influence of the atom-cavity coupling $\Omega$ on the squeezing factor $F_{1}$(black solid line) and the atomic excited population $P_{e}$(blue dotted-dashed line) versus $\gamma_{0}t$ in the Markovian regime($\lambda=5\gamma_{0}$). (a)$\Omega=\gamma_{0}$; (b)$\Omega=2\gamma_{0}$. Other parameters are $\theta=\frac{2\pi}{3}$, $\varphi=0$ and $\omega_{0}=10\gamma_{0}$. The red dashed lines indicate the envelope lines of the $F_{1}$-oscillations.}
\end{center}

In Figure 2, we plot the influence of the atom-cavity coupling $\Omega$ on the squeezing factor $F_{1}$ in the Markovian regime($\lambda=5\gamma_{0}$). From Figure 2, we can see that the $F_{1}$ is obviously dependent on the atom-cavity coupling. Figure 2(a) exhibits a quick oscillation of $F_{1}$ as time $t$ increases due to the interaction between the atom with the dissipative cavity. The squeezing first increases to $F_{1}=-0.13$ at $\gamma_{0}t=\pi/2$ from $F_{1}=0$ at $\gamma_{0}t=0$ and then collapses to zero at $\gamma_{0}t=\pi$. Afterwards, the squeezing is revived to $F_{1}=-0.03$ at $\gamma_{0}t=3\pi/2$, which is much smaller than $F_{1}=-0.13$. In a longer time scale, the oscillation of $F_{1}$ is modulated by the coupling parameter $\Omega$(see the envelope line expressed by the red dashed line in the Figure 2(a)). The collapse-revival of $F_{1}$-oscillation is obvious over $\gamma_{0}t\in[0,8]$ and the periods is $\pi/\Omega$. And this collapse-revival phenomenon is always accompanied by the decay and repopulation of the excited atom. The light field reaches its maximal squeezing when the $P_{e}$ reduces to zero from $0.25$. Subsequently, the $F_{1}$-oscillation collapses to zero when the $P_{e}$ again rises from zero. Finally, this collapse-revival of $F_{1}$-oscillations will disappear due to the dissipation of the cavities coupling with the Markovian environments. Comparing Figure 2(a) and (b), we can know that their squeezing dynamics is similar for different $\Omega$. The difference is in the collapse-revival frequency of $F_{1}$-oscillation and in the maximal squeezing obtained. The collapse-revival frequency of $F_{1}$-oscillation in the latter case is twice the former. The maximal squeezing obtained is obvious larger than the former. Hence, strengthening $\Omega$ can increase the collapse-revival frequency of $F_{1}$-oscillation and the maximal squeezing.

The physical interpretations of the above results are as follows. Squeezing light can be generated when the atom transits to a ground state from an excited state. And, due to the interaction between the atom and its cavity, the photons can be exchanged between the atom and its cavity so that the collapse-revival phenomenon of $F_{1}$-oscillation can occur in a short time. On the other hand, due to the coupling of the cavity with its bosonic environment, the photons exchanged to the cavity will continuously reduce thus the collapse-revival phenomenon of $F_{1}$-oscillation disappears in a long time, as shown in Figure 2(a)-(b).

\begin{center}
\includegraphics[width=16cm,height=14cm]{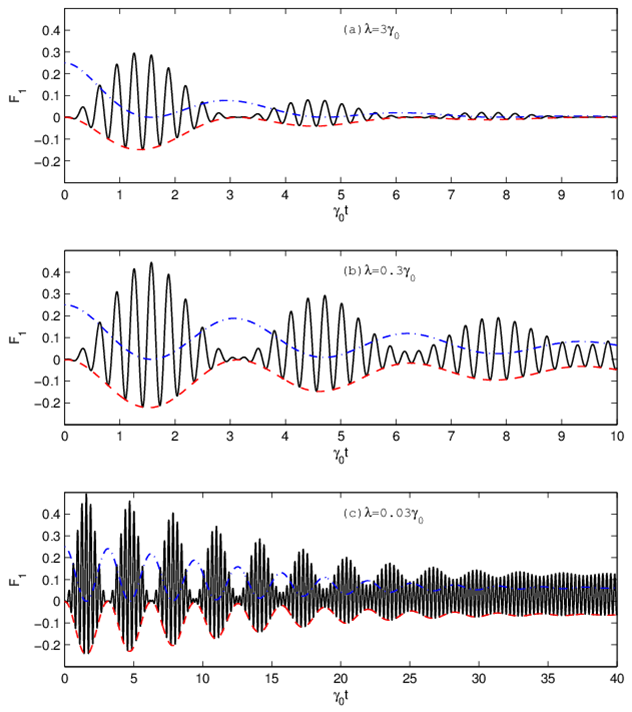}
\parbox{16cm}{\small{\bf Figure 3.}
(Color online)The influence of the non-Markovian effect on the squeezing factor $F_{1}$(black solid line) and the atomic excited population $P_{e}$(blue dotted-dashed line) versus $\gamma_{0}t$ when $\Omega=\gamma_{0}$. (a)$\lambda=3\gamma_{0}$; (b)$\lambda=0.3\gamma_{0}$; (c)$\lambda=0.03\gamma_{0}$. Other parameters are $\theta=\frac{2\pi}{3}$, $\varphi=0$ and $\omega_{0}=10\gamma_{0}$. The red dashed lines indicate the envelope lines of the $F_{1}$-oscillations.}
\end{center}

Figure 3 exhibits the influence of the non-Markovian effect on the squeezing factor $F_{1}$ when $\Omega=\gamma_{0}$. From Figure 3, we can find that, for the same $\Omega$, the collapse-revival frequencies of $F_{1}$-oscillations are same whether the Markovian or the non-Markovian regimes. The influence of the non-Markovian effect on the $F_{1}$ is embodied in the maximal value of squeezing. In the Markovian regime($\lambda=3\gamma_{0}$), the squeezing increases to $F_{1}=-0.14$ at $\gamma_{0}t=\pi/2$ from $F_{1}=0$ at $\gamma_{0}t=0$ and then collapses to zero at $\gamma_{0}t=\pi$. Afterwards, the squeezing is revived to $F_{1}=-0.04$ at $\gamma_{0}t=3\pi/2$, which is much smaller than $F_{1}=-0.14$. That is to say, the revival of $F_{1}$-oscillation is very close to zero due to the cavity dissipation, as shown in Figure 3(a). For Figure 3(b), when $\lambda=0.3\gamma_{0}$, due to the memory and feedback effect of the non-Markovian environment, the squeezing increases to $F_{1}=-0.21$ at $\gamma_{0}t=\pi/2$ from $F_{1}=0$ at $\gamma_{0}t=0$ and then collapses to zero at $\gamma_{0}t=\pi$. And the squeezing is revived to $F_{1}=-0.15$ at $\gamma_{0}t=3\pi/2$. This shows that the maximal value of squeezing in the non-Markovian regime is obviously better than the Markovian regime. In particulary, when $\lambda=0.03\gamma_{0}$, the maximum squeezing can reaches $F_{1}=-0.24$ in a short time and the collapse-revival phenomenon of $F_{1}$-oscillation is very remarkable. The $F_{1}$ will tend to a stably periodic oscillation between $-0.052$ to $0.124$ in a long time, as shown in Figure 3(c). Hence, the smaller the value of $\lambda$ is, the stronger the non-Markovian effect is, the more remarkable the collapse-revival phenomenon of $F_{1}$-oscillation is.

\begin{center}
\includegraphics[width=16cm,height=10cm]{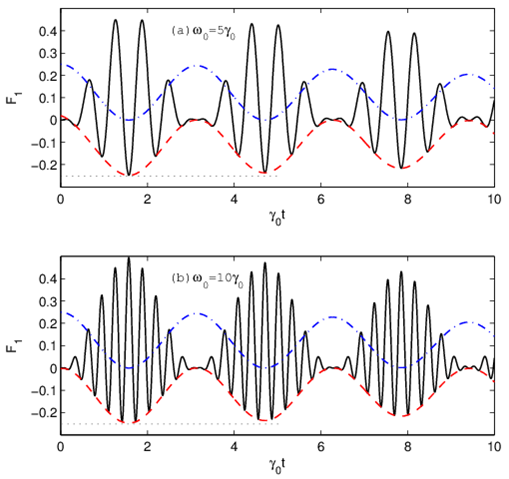}
\parbox{16cm}{\small{\bf Figure 4.}
(Color online)The influence of the atomic Bohr frequency $\omega_{0}$ on the squeezing factor $F_{1}$(black solid line) and the atomic excited population $P_{e}$(blue dotted-dashed line) versus $\gamma_{0}t$ when $\Omega=\gamma_{0}$. (a)$\omega_{0}=5\gamma_{0}$; (b)$\omega_{0}=10\gamma_{0}$. Other parameters are $\theta=\frac{2\pi}{3}$, $\varphi=0$ and $\lambda=0.01\gamma_{0}$. The red dashed lines indicate the envelope lines of the $F_{1}$-oscillations.}
\end{center}

The physical explanation is that the photons dissipated to the environment can be partly returned to the cavity field due to the memory and feedback effect of the non-Markovian environment. And the stronger the the non-Markovian effect is, the more the photons returned is. Therefor, in the Markovian or the weak non-Markovian regimes, the $F_{1}$ will decay to zero in a long time(see in Figure 3(a)-(b)). If the non-Markovian effect is very strong, the photons dissipated to the environment can be effectively fed back to the cavity so that the $F_{1}$-oscillation can be well revived after they decrease to zero(see Figure 3(c)).

Figure 4 displays the influence of the resonant frequency $\omega_{0}$ of the atom-cavity on the squeezing in the non-Markovian regime($\lambda=0.01\gamma_{0}$) and with $\Omega=\gamma_{0}$. From Figure 4, we know that the frequency of the $F_{1}$-oscillations is obvious dependent on the resonant frequency $\omega_{0}$, which arise from the resonant coupling of the atoms with its cavity. Comparing Figure 4(a) and (b), we find that the frequency of the $F_{1}$-oscillations in the latter case is twice the former because the resonant frequency in the latter case is twice the former. Thus, the more the value of $\omega_{0}$ is, the more violent the Rabi oscillating of the atom-cavity is.

\section{\fs Conclusion}
In conclusion, we investigate the squeezing of light field in the dissipative Jaynes-Cummings model by the time-convolutionless master-equation approach. We discuss in detail the influences of the initially atomic state, the atom-cavity coupling, the non-Markovian effect and the resonant frequency on the squeezing when $|\psi(0)\rangle=(cos(\frac{\theta}{2})|e\rangle+e^{i\varphi}sin(\frac{\theta}{2})
|g\rangle)_{A}|0\rangle_{f}$. The results show that, the atomic transition can generate squeezing light, and the collapse-revival of the $F_{1}$-oscillations can occur whether the Markovian or the non-Markovian regimes. The squeezing factor $F_{1}$ is obvious $\theta$ dependent but is $\varphi$ independent for different the initial atomic state and the maximal squeezing can be obtained when $\theta=\frac{2\pi}{3}$. The frequency of the $F_{1}$-oscillations is obvious dependent on the resonant frequency $\omega_{0}$, the more the value of $\omega_{0}$ is, the bigger the oscillating frequency of the squeezing is. Strengthening the atom-cavity coupling can increase the collapse-revival frequency of $F_{1}$-oscillation and the maximum squeezing obtained. The collapse-revival phenomenon of $F_{1}$-oscillation lies on the non-Markovian effect, the smaller the value of $\lambda$ is, the stronger the non-Markovian effect is, the more remarkable the collapse-revival phenomenon of $F_{1}$-oscillation is. These results may offer interesting perspectives for future applications of open quantum systems in quantum optical, microwave cavity QED implementations, quantum communication and quantum information processing.
\\

\textbf{Funding} \\
This work was supported by the Science and Technology Plan of Hunan Province, China(Grant no. 2010FJ3148) and the National Natural Science Foundation of China (Grant No.11374096).

\centerline{\hbox to 8cm{\hrulefill}}

\end{CJK*} %% 结束中文、日文、韩文使用环境
\end{document}